\documentclass[twocolumn,prb,superscriptaddress,showpacs]{revtex4-1}
\usepackage{amsmath,amssymb}
\usepackage{graphicx}
\usepackage{verbatim}
\usepackage{amsfonts}
\usepackage{dcolumn}
\usepackage{bm}
\usepackage{color}
\usepackage[colorlinks,citecolor=blue]{hyperref}

\begin{document}

\title{\large \bf Overcomplete Bases for $S=1$ Spin Liquids}

\author{Zheng-Xin Liu}
\email{liuzxhk@gmail.com}
\affiliation{Institute for Advanced Study, Tsinghua University, Beijing, 100084, P. R. China}

\begin{abstract}
For a $S=1$ system with even number of spins, the product states of two-body singlets, called the singlet pair states, are overcomplete bases for the Hilbert space of  many-body singlets. If the system contains odd number of spins, a singlet state can be decomposed as a superposition of all of the following configurations, in each configuration three of the spins form a three-body singlet and the remaining form two-body singlet pairs.  This indicates that $S=1$ spin liquids are essentially resonating-valence bond (RVB) states. Although generally this conclusion is no longer valid for $SO(3)$ symmetric $S>1$ systems, it can be generalized to integer spin-$S$ systems with enlarged $SO(2S+1)$ symmetry. Similar results can also be obtained for systems with $SU(n)$ symmetry. 
\end{abstract}

\maketitle

\section{Introduction}

A quantum spin liquid is a many-body singlet state without breaking the global spin rotation symmetry and the lattice symmetry. In two dimensions, spin liquids may support fractional excitations. A gapped spin liquid has degenerate ground states on a torus and carries topological order.\cite{Wen89, WenNiu90, HCJiangNP12} It was proposed that spin liquids may exist in geometrically frustrated spin-1/2 anti-ferromagnetic systems owning to strong quantum fluctuations.\cite{Anderson1987}  Searching for spin liquid states has attracted lots of research interest in condensed matter physics from both experimental\cite{Triangle03, Dmit08,SHLee07, THHan12} and theoretical sides.\cite{Motrunich05,JiangKagome08, Ran08, WhiteKagome11,DMRGKagome12,CSLKagome12,CSLSheng14}

Spin liquid states are also called resonating valence bond (RVB) states\cite{Anderson1987}, where a valence bond stands for a two-spin singlet just like a Cooper pair in $s$-wave superconductors and a RVB state is a superposition of all possible valence bond covering configurations. The validity of identifying a spin liquid state as a RVB state relies on a mathematical theorem, which says that the valence bond covering configurations, sometimes called singlet pair states (SPSs), are overcomplete to span the Hilbert space of many-body-singlet states.\cite{Caspers89} More precisely, if the system is arbitrarily divided into two subsystems A and B with equal number of spins, then the overcomplete bases can be restricted to the valence bond covering configurations where each valence bond  contains one spin from A and one from B. The RVB picture yields several practical approaches to construct spin liquid states, such as Gutzwiller projection of BCS-type mean field wavefunctions\cite{Gros88}, Liang-Doucot-Anderson variational method\cite{LiangDoucotAnderson88} and tensor network (or projected entangled pair states) construction.\cite{PEPsRVB12} These approaches have been widely used to study antiferromagnetic spin models and high temperature superconductors. 

On the other hand, spin-1 systems also have strong quantum fluctuations, and consequently spin liquid states may also exist in spin-1 frustrated anti-ferromagnetic systems. For example, 1-dimensional (1D) gapped $S=1$ spin liquid --- the Haldane phase,\cite{HaldanePLA1983, HaldanePRL1983} has attracted lots of research interest. Candidates of 2D $S=1$ spin liquids have also been reported in literature\cite{NiGa2S4_05, Cheng-Spin-1SL2011, JiangS1SL09,LiuZhouNg2011_Spin-1SL, Xu-S1SL2012,PALee12Gutz}. An interesting aspect for $S=1$ anti-ferromagnets is that there may exist non-Abelian spin liquids which support non-Abelian anyon excitations.\cite{Senthil_NonAbelian11}

There is a similarity between spin-1/2 systems and spin-1 systems, namely, $S=1/2$ is the fundamental representation of $SU(2)$ group whereas $S=1$ is the fundamental representation of $SO(3)$ group. So it is natural to extrapolate that $S=1$ spin liquid states are also RVB states, where a valence bond represents a singlet-pair formed by two $S=1$ spins.  In the remaining part of this paper, we show that this is indeed the truth. We prove that for $S=1$ systems, singlet pair states are also overcomplete to span the Hilbert space of many-body singlet states. Comparing with spin-1/2 systems, there are two differences: one is that here the number of spins of a many-body singlet can be either even or odd, the other is that the subsystem-singlet-pair states, where each singlet pair is from one subsystem to the other, are no longer complete bases. 

Above conclusion ensures that some methods used to study $S=1/2$ spin liquids can also be applied for $S=1$ systems. For instance, Gutzwiller projection approach based on fermionic slave particle representation provides very good trial wave functions for $S=1$ bilinear-biquadratic anti-ferromagnetic Heisenberg chains. \cite{LiuZhouTuWenNg2012, LiuZhouNgExt14}  It was shown that the Haldane phase is long-range RVB states (from this point of view, the resonating loop states\cite{YaoS1SL10} in 2D are also long-range RVB states) while the dimer phase is short-rang RVB states. Another  example is the tensor network approach which tells us that extremely short-ranged RVB state for $S=1$ spins on Kagome lattice carries $Z_2$ topological order.\cite{YaoS1SL10,RAL14,CXPollman14}

However, it should be noted that the RVB picture of spin liquid states is no longer valid for $SO(3)$ symmetric spin systems with spin magnitude $S>1$. In these large spin systems, spin-singlet-clusters (3-body-singlet, or 4-body-singlet, so on and so forth) as well as two-body singlet pairs are necessary to span the Hilbert space of many-body singlets.  The RVB representation will be valid for integer spin-$S$ systems\footnote{For half-odd-integer spin-$S$ systems, the $SO(3)$ symmetry can be enlarged into $SP(2S+1)$, since a $SO(3)$ singlet formed by two spin-$S$ spins is also a $SP(2S+1)$ singlet. The $SP(2S+1)$ symmetry is beyond the scope of the present paper and will be discussed in our future work.} when they have enhanced symmetry group $SO(2S+1)$.\cite{TuSO(n)13} 

Recently, systems with $SU(n)$ symmetry attracted lots of interest in cold atom physics,\cite{ScienceSU(n)14, NatPhysSU(n)24, SU(n)_Hubbard04, SU(n)_coldatom10} and $SU(n)$ spin liquids states have also been studied theoretically.\cite{HermeleSU(n)CSL09, TuSU(n)CSL14} To this end, we also discuss the complete bases for $S=1$ spin liquids with $SU(3)$ symmetry. Generalization of this result to $SU(n)$ systems is straightforward. 

The remaining part of the paper is organized as follows. In section \ref{sec:SO3}, we prove that singlet pair states are overcomplete bases for $S=1$ spin liquids with $SO(3)$ symmetry. A special case where the $S=1$ system has $SU(3)$ symmetry is discussed in section \ref{sec:SU3}. In section \ref{sec:S=1apply} we discuss a simple application of the $S=1$ RVB representation in 1D spin liquids.  Section \ref{sec:sum} is devoted to conclusions and discussions. 

\section{Overcompleten bases for $SO(3)$ symmetric $S=1$ spin liquids}\label{sec:SO3}

\subsection{Tensor Representation and Young Tableau}

Under $SO(3)$ operation, the three components of $S=1$ vary as a real vector. The three bases for $S=1$ can be combined into the familiar vector form
\begin{eqnarray*}
&&|x\rangle={1\over\sqrt2}(|-1\rangle-|1\rangle),\\
&&|y\rangle={i\over\sqrt2}(|-1\rangle+|1\rangle),\\
&&|z\rangle=|0\rangle.
\end{eqnarray*}
We denote these bases as $V^m=|m\rangle$ where $m=x,y,z$. Thus the Hilbert space of a system with $L$ spins  form a rank-$L$ reducible {\it real} tensor representation of $SO(3)$. 

Similar to the reduction of $SU(2)$ tensors (see Appendix \ref{app:SU2}), $SO(3)$ tensor representations can be reduced according to different representations of the permutation group of the tensor indices (\textit{i.e.}, the site indices). Different representations of the permutation group can be described by different Young diagrams.\cite{ChenJQBook04} Comparing to $SU(2)$ group, the complicity of $SO(3)$ group (and generally $SO(n)$ group) is that the same Young diagram stands for different $SO(3)$ representations. \footnote{For $SU(n)$ system, each Young diagram corresponds to an $SU(n)$ irreducible representation. However, for $SO(n)$ tensors, the representation space corresponding to a Young diagram is usually reducible, since the trace of two indices is invariant under $SO(n)$. So the same Young diagram may stand for a direct sum of different irreducible representations.} 
We will illustrate this issue by two simple cases: $L=2$ and $L=3$.

The direct product of two spins is represented as a real rank-2 tensor $T^{mn}=V_1^m\otimes V^n_2$. The reduction of the tensor contains three channels $1\otimes1=0\oplus1\oplus2$. The representations with total spin $S_t=0,2$ are symmetric under permutation of the two site indices, while the state with $S_t=1$ is antisymmetric for the site indices. As shown in Fig.~\ref{fig: 2site}, the permutation symmetries can be labeled by Young diagrams:

\begin{figure}[htbp]
\centering
\includegraphics[width=1.1in]{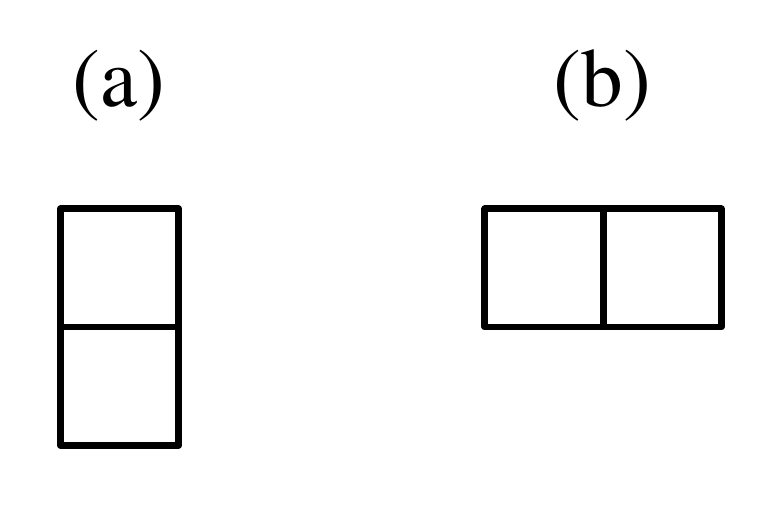}
\caption{Young diagrams for a 2-spin system. (a) The antisymmetric channel $[1;1]$, the total spin is $S_t=1$, which is called a dual vector; (b) The symmetric channel $[2]$, which contains the trace part $S_t=0$ and the traceless part $S_t=2$. } \label{fig: 2site}
\end{figure}

{\bf Antisymmetric channel} [Fig.~\ref{fig: 2site}(a), $S_t=1$]
\begin{eqnarray*}
T^{\{mn\}}=\sum_{mn}{1\over2}(T^{mn}-T^{nm}).
\end{eqnarray*}
Since there is only one free index, we can rewrite it as a dual vector:
\[
\tilde V^l=\sum_{mn}\varepsilon^{lmn}T^{\{mn\}},
\]
where $\varepsilon^{lmn}$ is the Levi-Civita symbol. In other words, the Young diagram $[1;1]$ is dual to the Young diagram $[1]$;

{\bf Symmetric channel} [Fig.~\ref{fig: 2site}(b), $S_t=0, 2$]

1) trace of $T$ ($S_t=0$)
\[\mathrm{Tr~}T=\sum_{mn}\delta^{mn}T^{mn}=\sum_m T^{mm};\]

2) traceless symmetric tensor ($S_t=2$)
\[T_0^{[mn]}=T^{[mn]}-{1\over3}\delta^{mn}\mathrm{Tr~} T.\]
where $T^{[mn]}={1\over2}(T^{mn}+T^{nm})$ and the subscript 0 in $T_0^{[mn]}$ means traceless, {\it i.e.} $\sum_{m,n}\delta^{mn}T_0^{[mn]}=0$.

\begin{figure}[htbp]
\centering
\includegraphics[width=1.8in]{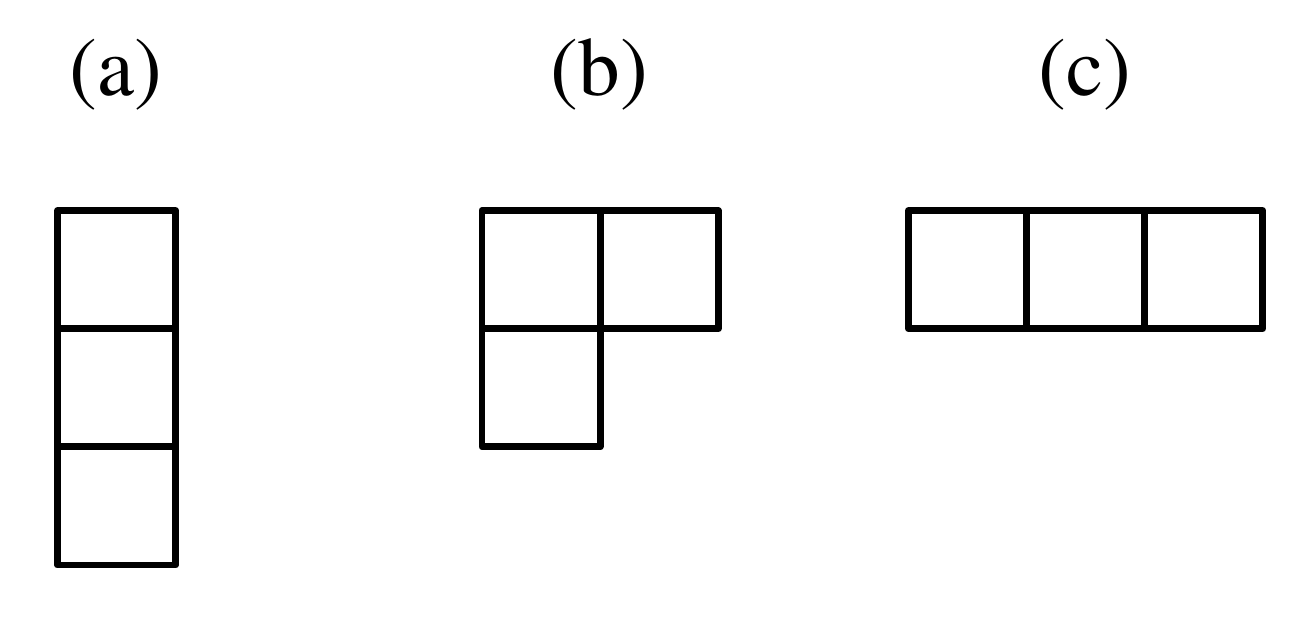}
\caption{Young diagrams for a 3-spin system. (a) The fully antisymmetric diagram $[1;1;1]$, $S_t=0$; (b) The mixed-symmetric diagram $[2;1]$, $S_t=1\oplus1\oplus2\oplus2$; (c) The fully symmetric diagram $[3]$, $S_t=1\oplus3$. } \label{fig: 3site}
\end{figure}

Similarly, the reduction of a rank-3 tensor can be labeled by Young diagrams shown in Fig.~\ref{fig: 3site}. The fully anti-symmetric diagram $[1;1;1]$ [see Fig. \ref{fig: 3site}(a)] stands for a singlet with $S_t=0$; The mixed symmetric diagram $[2;1]$ [see Fig.\ref{fig: 3site}(b)] stands for a direct sum of $S_t=1\oplus1\oplus2\oplus2$ (each irreducible representation is 2-fold degenerate because the Young diagram $[2;1]$ stands for a 2-dimensional representation of the permutation group); the fully symmetric diagram $[3]$ [see Fig.\ref{fig: 3site}(c)] represents a direct sum of $S_t=1\oplus3$. The bases for the irreducible representations are given in Appendix \ref{app: 1*3}. 

Above examples reveal important features for $S=1$ systems:
(1) three spins (or a rank-3 tensor) can form a singlet (or a scalar) if the indices are fully antisymmetric;
(2) a dual vector and a vector ({\it i.e.} a two-row unit and a one-row unit) cannot contract into a singlet;
(3) a rank-odd fully symmetric tensor can not form a singlet.

\begin{figure}[htbp]
\centering
\includegraphics[width=2.6in]{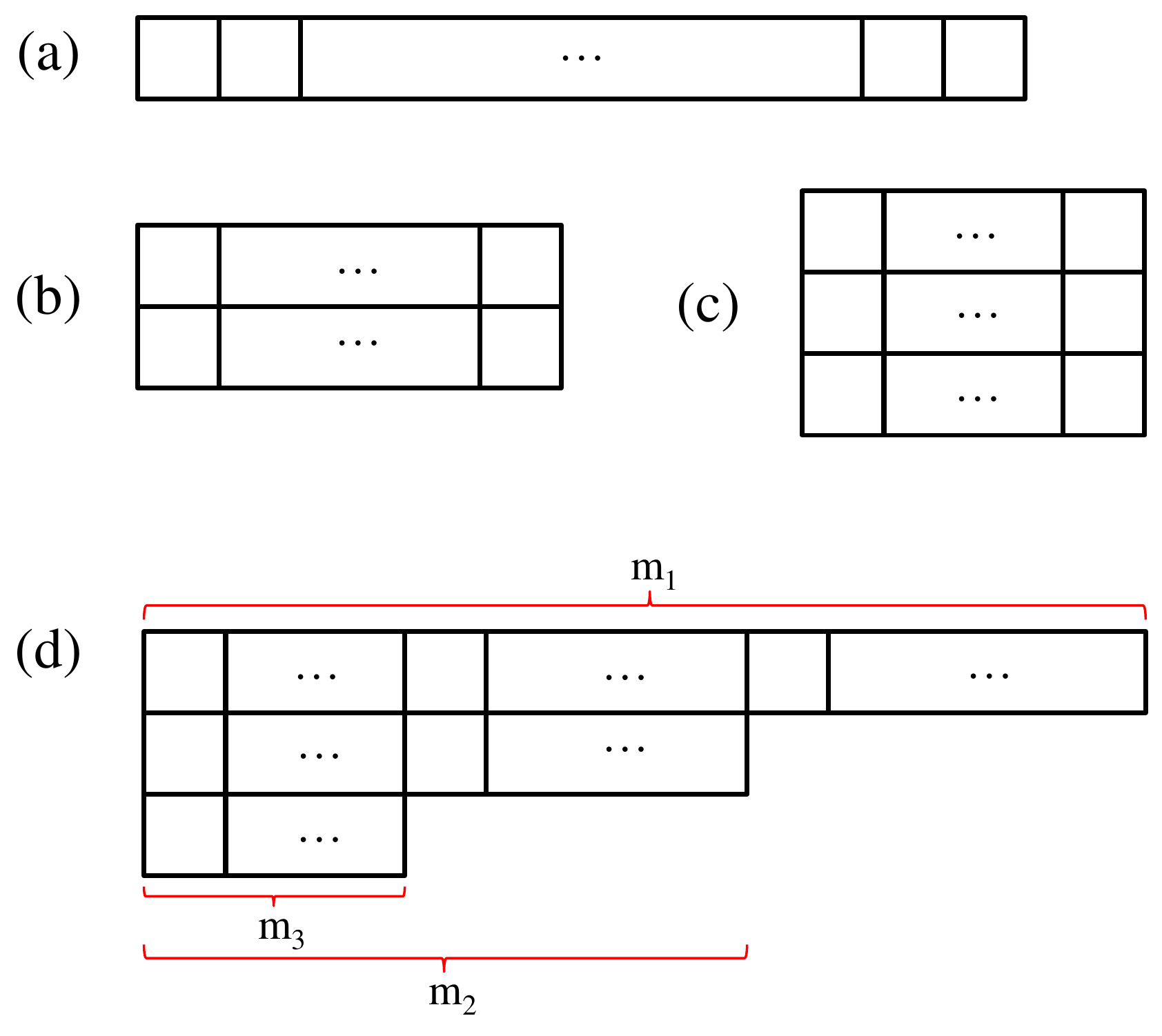}
\caption{The Young diagrams that contains singlet representations.  (a) the fully symmetric Young diagram $[2n_1]$; (b) the diagram $[2n_2;2n_2]$ is dual to the fully symmetric one, it contains singlet components if $N$=even; (c) three-row diagram $[n_3;n_3;n_3]$.  (d) The most general Young diagram $[m_1;m_2;m_3]$. Supposing the number of sites is $L$ ($L$ can be either even or odd), there are three conditions for the Young diagram if it contains a singlet: $m_1+m_2+m_2=L$; $(m_1-m_2)$=even; $(m_2-m_3)$=even.  } \label{fig: 3YT}
\end{figure}

According to above properties, not every Young diagram contains singlet representations. In the following we list the Young diagrams that contain singlet channels in their tensor reduction:

(A) A single row with an even number of columns,  
see Fig.~\ref{fig: 3YT}(a);

(B) Two equal-length rows with even number of columns,  
see Fig.~\ref{fig: 3YT}(b);

(C) Three equal-length rows, 
see Fig.~\ref{fig: 3YT}(c).

(D) A combination of above three, see Fig.~\ref{fig: 3YT}(d).

If a Young diagram contains singlet channels, it must be one of above cases. {\it The number of independent singlets corresponding to each Young diagram is equal to the dimension of the representation of the permutation group labeled by the same Young diagram} (see Appendix \ref{app:SU2}).

\subsection{Overcompleteness of SPSs}

Now we prove that all singlets corresponding to the diagrams in Fig.~\ref{fig: 3YT} can be expanded as superpositions of SPSs. 

\subsubsection{Two Formulas}

Firstly we give the following properties of fully antisymmetric tensor $\varepsilon^{abc}$,
\begin{eqnarray}
\varepsilon^{abc}\varepsilon^{def}&=&\delta^{ad}(\delta^{be}\delta^{cf}-\delta^{bf}\delta^{ce})
-\delta^{ae}(\delta^{bd}\delta^{cf}-\delta^{bf}\delta^{cd})\nonumber\\
&&-\delta^{af}(\delta^{be}\delta^{cd}-\delta^{bd}\delta^{ce}),\label{delt3}
\end{eqnarray}
specially, if $a=d$, Eq.~(\ref{delt3}) reduces to
\begin{eqnarray}
\sum_a\varepsilon^{abc}\varepsilon^{aef}=\delta^{be}\delta^{cf}-\delta^{bf}\delta^{ce}.\label{delt2}
\end{eqnarray}

Above two are the most important equations in proving the completeness of SPSs. 

\subsubsection{Simple Applications}

Now we illustrate two applications of these formulas:

\begin{figure}[htbp]
\centering
\includegraphics[width=1.2in]{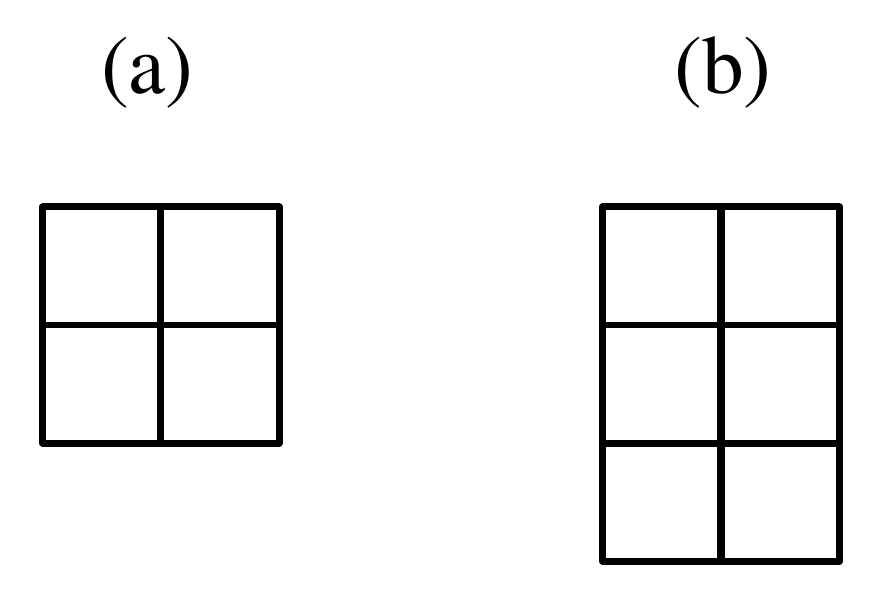}
\caption{The singlets described by the following Young diagrams can be expanded as superpositions of SPSs: (a) 4 spins with permutation symmetry channel described by the diagram $[2;2]$; (b) 6 spins with symmetry channel $[2;2;2]$. } \label{fig: 4_6site}
\end{figure}

(1) The Young diagram $[2;2]$ in Fig.\ref{fig: 4_6site}(a) is dual to Fig.\ref{fig: 2site}(b). This diagram contains singlet channels, meaning that two dual vectors can contracte into a singlet. From Eq.~(\ref{delt2}), we have
\begin{eqnarray}\label{T22}
|0,0\rangle&=&\sum_{abcef}T_{[2;2]}^{bcef}\varepsilon^{abc}\varepsilon^{aef}\nonumber\\
&=&\sum_{bc}(T^{bcbc}-T^{bccb}) 
\end{eqnarray}
here the indices of $T_{[2;2]}^{bcef}$ respect the symmetry described by Young diagram $[2;2]$, namely, $\{bc\}$ and $\{ef\}$ are anti-symmetric and then $[be]$ and $[cf]$ are symmetric under permutation. Above equation means that the 4-spin singlet described by $[2;2]$ can be expanded as a superposition of SPSs $|0,0\rangle=(13)(24)-(14)(23)$, where the bracket means a singlet pair.

(2) The Young diagram $[2;2;2]$ in Fig.\ref{fig: 4_6site}(b) contains singlet component. Owning to the relation (\ref{delt3}) we have
\begin{eqnarray}\label{T222}
|0,0\rangle&=&\sum_{abcdef}T_{[2;2;2]}^{abcdef}\varepsilon^{abc}\varepsilon^{def}\nonumber\\
&=&\sum_{abc}[(T^{abcabc}-T^{abcacb})-(T^{abcbac}-T^{abccab})\nonumber\\
&&-(T^{abccba}-T^{abcbca})].
\end{eqnarray}
This means that the 6-spin singlet described by $[2;2;2]$ can be written as a superposition of SPSs $|0,0\rangle=[(14)(25)(36)-(14)(26)(35)]-[(15)(24)(36)-(15)(26)(34)]-[(16)(25)(34)-(16)(24)(35)]$.

\subsubsection{Proof of Overcomplete Bases}

With these pre-knowledge, we are ready to prove the overcompleteness of SPSs. Firstly we suppose that the number of sites is even $L=2N$. In each SPS there are $N$ singlet pairs and the total number of SPSs is equal to ${(2N)!\over(N!)(2^N)}$. Now we show that the all of Young diagrams  listed in Fig.~\ref{fig: 3YT} can be expanded by these SPSs.

(A) The single-row Young diagram $[2N]$ 

The Hilbert space of $2N$ spins can be described by a rank-$2N$ real tensor $T^{m_1m_2...m_{2N}}$. The Young diagram $[2N]$ means these indices are fully symmetric under permutation, namely, 
\[
T^{m_1m_2...m_{2N}}_{[2N]} ={1\over (2N)!}\sum_{ P}T^{P(m_1m_2...m_{2N})},\] 
where $P$ means permutation of the $2N$ indices $(m_1,m_2,...,m_{2N})$. To obtain a singlet, all indices should be contracted two by two:
\[
|0,0\rangle= \sum_{\{m_i\}}\prod_{k=1}^N \delta^{m_{i_k}m_{j_k}}T^{m_1m_2...m_{2N}}_{[2N]},
\]
where the two spins at sites $i_k$ and $j_k$ form a singlet pair, and $|0,0\rangle$ is an equal weight superposition of all SPSs.

(B) The two-row Young diagram  $[N;N]$ contains singlet representations if $N$ is even. Since $[N;N]$ is dual to $[N]$, we can treat the two spins at each column as a dual vector, and the argument in (A) still works for the dual vectors. Then we can use relation (\ref{delt2}) to express the singlet formed by dual vectors as SPSs [referring to Eq.(\ref{T22})];

(C) The three-row Young diagram  $[M;M;M]$ [This Young diagram is available if $L$ is divisible by 3, namely, $L=3M$ ($M$ is even since $L$ is even). Otherwise this diagram can only occur as part of the Young diagram in Fig.~\ref{fig: 3YT}(d)]. The three spins in each column form a singlet. Similar to Eq.~(\ref{T222}), the product of three-spin-singlets can be expressed as superposition of SPSs using relation (\ref{delt3});

(D) A general Young diagram $[m_1;m_2;m_3]$ is a combination of above three diagrams with $m_1+m_2+m_3=L$ and $(m_1-m_2)=$even, $(m_2-m_3)=$even. If $L$=even, then $m_3$=even, otherwise $m_3$=odd. The three-row part $[m_3;m_3]$ is already an singlet. The two-row part $[(m_2-m_3);(m_2-m_3)]$ can not contract with the one-row part $[m_1-m_2]$ owning to our previous argument [see Fig. \ref{fig: 3site}(b)]. So if the whole diagram contains singlet channels, then the two-row part and the one-row part must form singlets independently. Thus, all the three parts in the combined diagram can be considered independently, and the arguments in (A),(B),(C) still work. As a result, the singlet states described by a general diagram $[m_1;m_2;m_3]$ can be written as superpositions of SPSs.

Now we consider the case when $L=$odd, where the system can not be completely grouped into singlet pairs. However, if we arbitrarily select out three spins to form a three-spin singlet, then the remaining spins can completely combine into singlet pairs. So possibly any singlet of the system can be expanded as a superposition of all of the following configurations, in each configuration three spins form a singlet and the remaining form singlet pairs. Now we show how this is true. Since $L$=odd, (A) and (B) are not relevant. The Young diagram (C) is relevant if $L$ is divisible by 3, namely, $L=3M$ where $M$ is odd, and if a Young diagram in case (D) contains singlets, $m_3$ must be odd. In both cases, if we single out the first column (which corresponds to three arbitrary spins), then the remaining part can be expanded as SPSs according to our previous discussion. As a result, the singlet with odd number of spins can be decomposed as a superposition of products of a three-spin-singlet and singlet pairs. 

\subsubsection{Check The Completeness of Young Diagrams}
We denote the Hilbert space spanned by the singlets of a $L$-spin system as $\mathcal H_0$. It is not difficult to see that the dimension of the Hilbert space $\mathcal H_0$ is equal to the difference between the number of states with $S_z=0$ and the number of states with $S_z=1$:
\begin{eqnarray}\label{dimH0}
d_{\mathcal H_0,L}=\sum_{i=0}^{\{{L\over2}\}} C_{L}^iC_{L-i}^i-\sum_{i=1}^{\{{L+1\over2}\}} C_{L}^iC_{L-i}^{i-1},
\end{eqnarray}
where $C_L^i={L!\over i!(L-i)!}$, and $\{{L\over2}\}$ is the integer part of $L/2$. 

To see if the bases described by the Young diagrams are exhausting, we can compare (\ref{dimH0}) with the sum of the dimensions of all permitted Young diagrams (refer to Appendix \ref{app:SU2}) listed in (A)$\sim$(D).  The consistency can be checked for small $L$. For example, when $L=7$, above formula gives $d_{\mathcal H_0,7}=36$, and the two possible Young diagrams give $d_{[5;1;1]}+d_{[3;3;1]}=15+21=36$; when $L=8$, above formula gives $d_{\mathcal H_0,8}=91$, in consistent with the result by summing all possible Young diagrams $d_{[4;2;2]}+d_{[4;4]}+d_{[6;2]}+d_{[8]}=56+14+20+1=91$.

Above we verified that the Young diagrams in Fig.\ref{fig: 3YT} include all states of $\mathcal H_0$. We have also shown that the singlet states corresponding to every Young diagram in Fig.\ref{fig: 3YT} can be destructed into superposition of products of 2-body singlets (and a 3-body singlet if $L$=odd).  Synthesizing these two points we conclude that every singlet of the system can be written in forms of superposition of products of 2-body singlets (and a 3-body singlet if $L$=odd). This finishes the proof of overcomplete bases for many-body singlets of $SO(3)$ symmetric $S=1$ systems.

\section{overcomplete bases for $SU(3)$ symmetric systems}\label{sec:SU3}

With particular interactions, $S=1$ models may have an enhanced $SU(3)$ symmetry. For example, the $J$-$K$ model 
\begin{eqnarray}\label{JK}
H=\sum_{\langle i,j\rangle}[J\mathbf S_i\cdot\mathbf S_j + K(\mathbf S_i\cdot\mathbf S_j)^2]
\end{eqnarray}
with $J=K$ is invariant under $SU(3)$ and now $S=1$ carries the fundamental representation of $SU(3)$ group. For this kind of $SU(3)$ systems, a singlet unit contains at least three spins. We will show that if the ground state of a many-body system does not break $SU(3)$ symmetry (if lattice symmetry is unbroken, then it is a $SU(3)$ spin liquid), then it can be expanded as superposition of products of three-body-singlet-clusters, called  singlet-cluster states (SCSs).

Similar to $SU(2)$ systems, an irreducible representation of $SU(3)$ can also be uniquely labeled by a Young diagram(see Appendix \ref{app:SU2}). $SU(3)$ singlets are described by the Young diagram of three rows with equal number of columns [see Fig.~\ref{fig: 3YT}(c)]. That is to say, if the ground state is a $SU(3)$-singlet, then the system size must be divisible by 3, say, $L=3M$. The dimension of the Hilbert space of singlets is equal to the dimension of the $[M;M;M]$ representation of the permutation group, which is equal to $d_{\mathcal H_0,3M}=d_{[M;M;M]}={2(3M)!\over (M+2)!(M+1)!M!}$ (see Appendix \ref{app:SU2}).

The total number of all possible SCSs is ${(3M)!\over (M!)6^M}$. It is easy to see that these bases are overcomplete for $\mathcal H_0$. However, the number of overcomplete bases can be significantly reduced. To see this, we arbitrarily divide the $3M$-site system into three subsystems, each containing $M$ sites. If we require that the three spins in each singlet cluster come from three different subsystems, then the total number of subsystem-SCSs becomes $(M!)^2$. Following the proof of overcompleteness of subsystem-SPSs for $SU(2)$ systems(see Appendix \ref{app:SU2}), one can show that any $SU(3)$ singlet can be expanded as a superposition of these $(M!)^2$ subsystem-SCSs.

\section{Application in Gutzwiller approach of excited states}\label{sec:S=1apply}  

In this section, we will apply the SPS bases to the excited states of 1D Haldane phase [namely, the model (\ref{JK}) with $-1<K/J<1,\ J>0$] in the Gutzwiller approach, and prove that the one-magnon excited states and two-magnon excited states obtained are orthogonal.  Noticing that a single magnon carries spin-1, so the total spin of two magnons can be 0,1 or 2. The orthogonality between a one-magnon state and a two-magnon state is obvious if they carry different spin angular momentum or lattice momentum. In the following we will show that they are still orthogonal even if the two states carry the same quantum numbers.

We first briefly review the Gutzwiller approach for the Haldane phase.\cite{LiuZhouTuWenNg2012, LiuZhouNgExt14} The Gutzwiller approach for $S=1$ spin models is based on the fermion representation of spin-1 spins, where three species of fermions (called spinons) $f_x, f_y, f_z$ are introduced to rewrite the spin operators as 
\[S^\alpha_i=\sum_{\beta,\gamma}i\varepsilon^{\alpha\beta\gamma}f_{\beta,i}^\dag f_{\gamma,i}, \ \  {\rm with\ } \alpha, \beta, \gamma = x,y,z\] 
under an onsite particle number constraint 
\begin{eqnarray}\label{N=1}
f_{x,i}^\dag f_{x,i}+f_{y,i}^\dag f_{y,i}+f_{z,i}^\dag f_{z,i}=1.
\end{eqnarray}
In this fermion representation, the spin model (\ref{JK}) is rewritten as $H=-\sum_{\alpha,\beta,\langle i,j\rangle}[Jc_{\alpha,i}^\dag c_{\alpha,j} c_{\beta,j}^\dag c_{\beta,i}+(J-K)c_{\alpha,i}^\dag c_{\alpha,j}^\dag c_{\beta,j}c_{\beta,i}]$, and its ground state and low energy excited states can be approximately described by Gutzwiller projected eigenstates of the following mean field Hamiltonian: 
\begin{eqnarray}\label{Hmf}
H_{\rm mf} &=& \sum_{\alpha,\langle i,j\rangle} (\chi f_{\alpha,i}^\dag f_{\alpha,j}+\Delta f_{\alpha,i}^\dag f_{\alpha,j}^\dag + {\rm h.c}) +\sum_{\alpha,i} \lambda f_{\alpha,i}^\dag f_{\alpha,i}\nonumber\\
&=& \sum_{\alpha,k} E_k \Gamma_{\alpha,k}^\dag \Gamma_{\alpha,k},
\end{eqnarray}
where $\Gamma_{\alpha,k}$ are Bogoliubov particles and $\chi, \Delta, \lambda$ are variational parameters determined by minimizing the trial ground state energy $E_{\rm Grd} = \langle{\rm Grd}|H|{\rm Grd}\rangle/\langle{\rm Grd}|{\rm Grd}\rangle$ with $|{\rm Grd}\rangle =P_G |{\rm mf(\chi, \Delta, \lambda)}\rangle$. Here $P_G$ means Gutzwiller projection that enforces the constraint (\ref{N=1}).  When $|\lambda|<2|\chi|$ and $\Delta\neq0$, above mean field model describes a topological superconductor and the Gutzwiller projected ground state belongs to the Haldane phase.  

A subtle property of topological superconductor is that the fermion parity of its ground state depends on boundary condition.\cite{Kitaev2001, LiuZhouTuWenNg2012} Without loss of generality, we assume that the length $L$ of the chain is even, then the fermion parity of the ground state is even under anti-periodic boundary condition and is odd under periodic boundary condition. As a consequence, we need to carefully choose boundary conditions to construct the low energy states of the Haldane phase. For example, the ground state of Haldane phase is given by
\[
|{\rm Grd}\rangle =P_G |{\rm mf}\rangle_{\rm apbc},
\]
where $|{\rm mf}\rangle_{\rm apbc}$ is the ground state of (\ref{Hmf}) under anti-periodic boundary condition. 
 
When obtaining one-magnon excited states, we should choose periodic boundary condition,
\[
| (\alpha,k+\pi)_{\rm 1-mag}\rangle =P_G \Gamma_{\alpha,k}^\dag |{\rm mf}\rangle_{\rm pbc},
\] 
where $k={2n\pi\over L},\ n=-{L\over2},-{L\over2}+1,...,{L\over2}-1$ and the extra momentum $\pi$ is owning to the change of boundary condition. Since the pairing term vanishes at momentum $k=0$, the three fermions $f^x_{k=0}, f^y_{k=0}, f^z_{k=0}$ are unpaired and they form a singlet. Except for these three spinons and the excited magnon $\Gamma_{\alpha,k}$ (when $k\neq0$ the magnon is essentially a broken Cooper pair with one spinon removed), the remaining spinons form Cooper pairs. Noticing that there is a three-body singlet (except for the case when the momentum of the 1-magnon excited state is $\pi$), the Young diagrams describing the one-magnon states have the shape shown in Fig.~\ref{fig: 3YT}d with $m_3=$odd, $m_2-m_3=$even, $m_1-m_2$=odd. 

For two-magnon excited states, we should use anti-periodic boundary condition. The  states with total spin-1 is given as
\[
|(\alpha,k_1+k_2)_{\rm 2-mag} \rangle =P_G \sum_{\beta,\gamma}\varepsilon^{\alpha\beta\gamma}\Gamma_{\beta,k_1}^\dag \Gamma_{\gamma,k_2}^\dag |{\rm mf}\rangle_{\rm apbc},
\] 
where $k_1,k_2 = {2n\pi\over L }+{\pi\over L},\ n=-{L\over2},...,{L\over2}-1$. In this case, all of the spinons  form Cooper pairs except for the two anti-symmetric magnons $\Gamma_{\beta,k_1}$ and $\Gamma_{\gamma,k_2}$. The Young diagrams describing the two-magnon states have the shape shown in Fig.~\ref{fig: 3YT}d with $m_3=$even, $m_2-m_3=$odd, $m_1-m_2$=even.

Since an one-magnon state and a two-magnon state are described by different Young diagrams, they must be orthogonal to each other. This result has been verified numerically. Generally, an odd-magnon excited state and an even-magnon state are orthogonal. 

Although the one-magnon state $|(\alpha,k)_{\rm 1-mag}\rangle$ and the two-magnon state $|(\alpha,k)_{\rm 2-mag}\rangle$ are orthogonal, they are not eigenstates of the spin Hamiltonian $H$, and the off-diagonal matrix elements $\langle(\alpha,k)_{\rm 1-mag}|H|(\alpha,k)_{\rm 2-mag} \rangle$ are usually nonzero. The off-diagonal entries will `mix' the two-magnon states with the one magnon state. The mixing will be remarkable  if the diagonal terms are equal $\langle (\alpha,k)_{\rm 1-mag}|H|(\alpha,k)_{\rm 1-mag}\rangle=\langle(\alpha,k)_{\rm 2-mag}|H|(\alpha,k)_{\rm 2-mag}\rangle$, in this case the one-magnon excitation will be unstable and will decay into two magnons. \cite{Magnondecay12, LiuZhouNgExt14}  For this reason, the one-magnon excitations, which are well defined around the momentum $k=\pi$ (since there is a finite gap between the one-magnon state and the two-magnon state with the same momentum), will become unstable in the vicinity near $k=0$ because the single-magnon dispersion merges into the two-magnon continuum.

\section{conclusion and discussion}\label{sec:sum}

In summary, we have shown that for a $S=1$ system with $L$ spins, every singlet of the system can be written in forms of a superposition of singlet-pair-states if $L=$even, or a superposition of 2-body-singlet pairs time a 3-body-singlet if $L$=odd. We have also proved that if the system has $SU(3)$ symmetry, the products of 3-body singlet states are overcomplete bases of many-body $SU(3)$ singlets. Our conclusion provides solid foundations for generalizing the methods used in $S=1/2$ resonating valence bond states to study $S=1$ systems. As a simple application, we showed that in the Gutzwiller approach the one-magnon excited states and two-magnon excited states are orthogonal even if they carry the same quantum numbers.

Our conclusion for $SO(3)$ symmetric spin-1 systems can be straightforwardly generalized to $SO(2n+1)$ symmetric $S=n$ systems if $n$ is an integer, namely, a $SO(2n+1)$ singlet can be written in forms of a superposition of singlet-pair-states if $L=$even, or a superposition of 2-body-singlet pairs time a $(2n+1)$-body-singlet if $L$=odd. This conclusion can also be generalized to $SO(2n)$ systems, where an $SO(2n)$ singlet contains even number of objects, and the overcomplete bases include all of the following states: (1) product of 2-body singlet pairs; (2) product of a $2n$-body-singlet times 2-body singlet pairs. However, the $SO(2n)$ symmetry CANNOT emerge in an $SO(3)$ symmetric spin-$(n-{1\over2})$ system since an $SO(2n)$ singlet may not be invariant under $SO(3)$. For example, the 2-body $SO(2n)$ singlet (which is symmetric under exchanging of the two objects) is different from the $SO(3)$ singlet formed by two spins (which is anti-symmetric under exchanging the spins), namely, the 2-body $SO(2n)$ singlet is NOT invariant under $SO(3)$ spin rotation. This means that $SO(2n)$ systems are very different from the usual spin systems. Finally, our conclusion for $SU(3)$ systems can be generalized to $SU(n)$ systems if the physical degrees of freedom carry fundamental representation of $SU(n)$ [notice that the $SO(3)$ symmetry for a spin-$({n-1\over2})$ system can be enlarged into $SU(n)$]. 


The author thank Hong-Hao Tu for encouraging him writing out this article and some valuable discussions as well as comments to the manusrcript. We also thank Zhong-Qi Ma, Fa Wang, Xiong-Jun Liu and Jason Ho for helpful dissuasions, and thank Tai-Kai Ng, Yi Zhou for previous collaborations. This work was initiated in IAS of HKUST during a program in 2013. We thank the support from NSFC 11204149 and Tsinghua University Initiative Scientific Research Program.

\appendix

\section{Overcompleteness of singlet pair states for spin-1/2 system}\label{app:SU2}

In this appendix, we will illustrate that, for a $2N$-site system with $SU(2)$ symmetry, singlet pair states (SPSs) are overcomplete to span  $\mathcal H_0$---the total Hilbert space for $2N$-body-singlets.

Before proving the completeness of SPSs for $SU(2)$, we briefly review the irreducible representation theory of $SU(n)$ group. All irreducible representations of $SU(n)$ can be reduced from $SU(n)$ tensors. For a fixed number of sites (which is equal to the rank of the tensor), the reduction of the tensor will fall back on the permutation group $P$, which is formed by permutations of the site indices of the tensor. Since $SU(n)$ and $P$ commute, they have common representation spaces labeled by Young diagrams. Since each Young diagram uniquely labels a representation space for $SU(n)$ and $P$, when we talk about a representation we can just refer to the corresponding Young diagram.

The bases of a $SU(n)$ tensor can be symbolized by Young tableaus --  Young diagrams filled with spin and site indices. In a Young tableau, every box is occupied with a spin variable and a site index, where each site number occurs only once. The spin indices in a Young tableau respect the following symmetry: the indices in the same column are anti-symmetrized, and then the indices in the same row are symmetrized. A Young tableau is called a standard Young tableau if the site indices have the following order: in each column the a index is bigger than the ones above it, and in each row the a index is bigger then the ones on its left. The number of ways to fill the site indices into a standard Young tableau is equal to the dimension of representation for the permutation group labeled the Young diagram, which can be simply calculated by hook numbers.\footnote{The dimensionality of a representatoin for permutation group characterized by a Young tabel can be simply calculated using the hook number. See \href{http://en.wikipedia.org/wiki/Young_tableau}{Wikipedia: Young tableau}}
Fixing the site configuration and varying the spin indices, one obtains the bases for a irreducible representation of $SU(n)$.

\begin{figure}[htbp]
\centering
\includegraphics[width=1.8in]{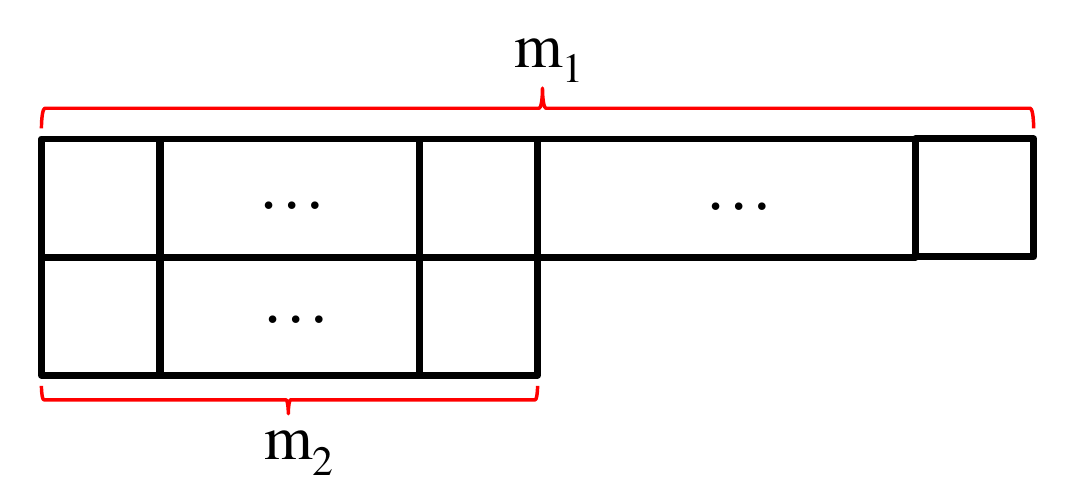}
\caption{A Young diagram $[m_1;m_2]$, $m_1+m_2=2N$. The total spin of this diagram is $S_t={1\over2}(m_1-m_2)$. } \label{fig: m1m2}
\end{figure}

\begin{figure}[htbp]
\centering
\includegraphics[width=2.4in]{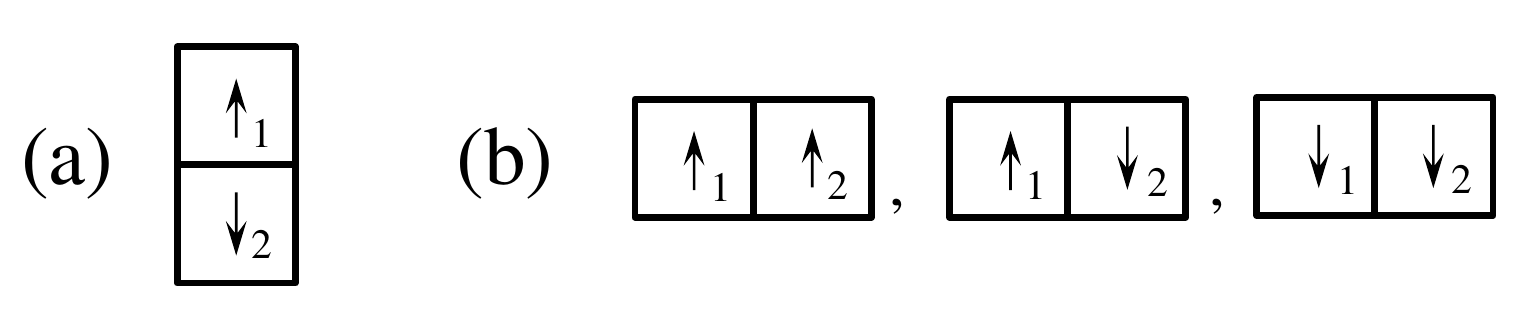}
\caption{The Young tableaus (labeling irreducible bases) for two spins. Different Young diagrams stand for different representations, and the bases in each representation are characterized by the corresponding Young tableaus. (a) basis for the spin singlet $S_t=0$ representation; (b) bases for the spin triplet $S_t=1$ representation. The three bases span a 3-dimensional representation of $SU(2)$ and 3-fold degenerate 1-dimensional representations of $P$.} \label{fig: 2spin}
\end{figure}

For $SU(2)$, an allowed Young diagram (and the corresponding Young tableau) contains at most two rows. As shown in Fig.\ref{fig: m1m2}, the total spin is equal to $S_t=(m_1-m_2)/2$, where $m_1$($m_2$) is the number of columns of the first(second) row. For example, for a two-spin system, there are two Young diagrams, one stands for a singlet the other represents a triplet. The Young diagrams and the corresponding Young tableaus are given in Fig.\ref{fig: 2spin}.

\begin{figure}[b]
\centering
\includegraphics[width=2.8in]{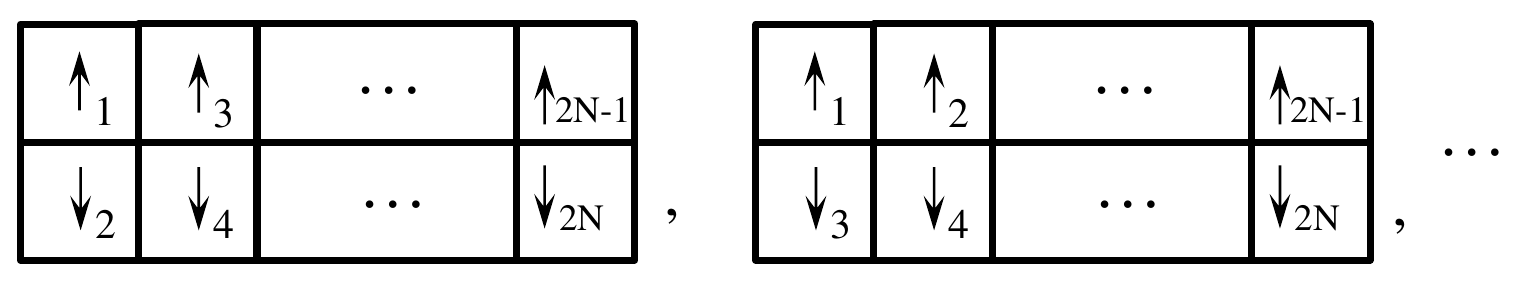}
\caption{The bases for $2N$-body singlet.} \label{fig: H0base}
\end{figure}

The Young diagram for the representation $S_t=0$ has two rows with $m_1=m_2=N$ (we will refer to this Young diagram as $[N;N]$). This Young diagram also stands for a $d_{[N;N]}={(2N)!\over N!(N+1)!}$ dimensional representation of $P$, where the $d_{[N;N]}$ bases are labeled by $d_{[N;N]}$ standard Young tableaus (corresponding to $d_{[N;N]}$ different site-index configurations). Since each Young tableau stands for a spin singlet, the spin indices can be fixed by filling $\uparrow$ in the upper boxes and $\downarrow$ in the lower ones (switching a pair of them results in a minus sign), see Fig.\ref{fig: H0base}. 
The bases corresponding to these Young tableaus are linearly independent (but not necessarily orthogonal). As a consequence, we obtain $d_{[N;N]}$ linearly independent singlet states. These states form complete bases for the Hilbert space of $\mathcal H_0$.

Now we prove the overcompleteness of SPSs as basses of $\mathcal H_0$. Since the two spins at the same column are antisymmetric and form a singlet, obviously the $d_{[N;N]}$ standard Young tableaus are superpositions of SPSs. Here the total number of these SPSs is ${(2N)!\over N!2^N}\gg d_{[N;N]}$. It is obvious that these SPSs are overcomplete for the Hilbert space $\mathcal H_0$. In the following we will show that the number of SPSs can be reduced without affecting the over-completeness. To this end, we arbitrarily divide the system into two subsystmes $A$ and $B$, each one containing $N$ spins ({\it e.g.}, $A$ and $B$ can be chosen as two sublattices for a bipartite lattice). If we require that every singlet pair in each SPS is formed by two spins coming from different subsystems, then the number of SPSs is reduced to $N!$. It turns out that these $N!$ SPSs are still overcomplete.

To verify this result, we notice that in a many-body singlet state, a spin $a_1$ in subsystem $A$ can not be symmetrized with all the spins in subsystem $B$, because this can be done only in a Young diagram with more than $N$ columns, which no longer contains $S_t=0$ representations. So $a_1$ must be antisymmetric with one of the spins in $B$. That is to say, any $S_t=0$ state can be written as a superposition of the following states:
\begin{eqnarray}
|0,0\rangle=\sum_{b_{i_1}}|\{a_1,b_{i_1}\}\rangle\times|(A',B')\rangle_{b_{i_1}},
\end{eqnarray}
where $|\{a_1,b_{i_1}\}\rangle$ means that $a_1$ and $b_{i_1}$ are antisymmetric under exchange site index which results in a singlet. $|(A',B')\rangle_{b_{i_1}}$ is the state formed by the remaining $(2N-2)$ spins in $A$ and $B$ and is also a singlet.

Since $|(A',B')\rangle_{b_{i_1}}$ is a $(2N-2)$-site singlet, we can repeat the above argument to find a partner in $B'$ for $a_2$. Repeating this procedure once and once again, we finally obtain:
\begin{eqnarray}
|0,0\rangle=\sum_{\{b_{i}\}}|\{a_1,b_{i_1}\}\rangle|\{a_2,b_{i_2}\}\rangle...|\{a_N,b_{i_N}\}\rangle,
\end{eqnarray}
Thus the over-completeness of the $N!$ SPSs is proved.

\section{Irreducible bases for three $S=1$ spins}\label{app: 1*3}

\begin{figure}[htbp]
\centering
\includegraphics[width=2.in]{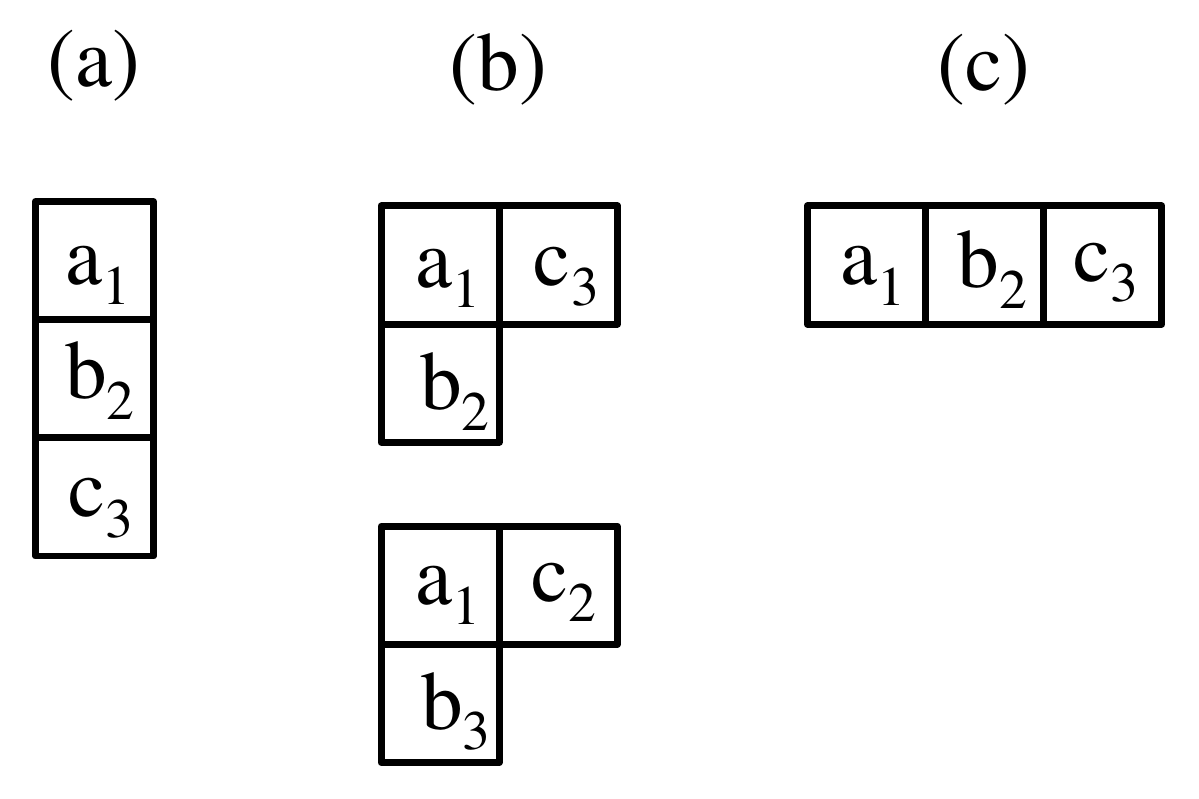}
\caption{Young diagrams for a 3-spin system. (a) The fully antisymmetric diagram $[1^3]$, $S_t=0$; (b) The mixed-symmetric diagram $[21]$, $S_t=1\oplus1\oplus2\oplus2$; (c) The fully symmetric diagram $[3]$, $S_t=1\oplus3$. } \label{fig: 3site_bases}
\end{figure}

For a rank-3 tensor, the irreducible representations can be classified by three Young diagrams. The diagram in Fig. \ref{fig: 3site_bases}(a) stands for a singlet $S_t=0$,
\[|0,0\rangle=T^{\{abc\}}=\sum_{abc}\varepsilon^{abc}T^{abc},
\]
namely $|0,0\rangle\propto T^{xyz}-T^{xzy}-T^{yxz}-T^{zyx}+T^{zxy}+T^{yzx}$.

Fig.\ref{fig: 3site_bases}(b) stands for a direct sum of $S_t=1\oplus1\oplus2\oplus2$. The bases for $S_t=1$ are
\begin{eqnarray*}
V^m&=&\sum_{nabc}\varepsilon^{mnc}\varepsilon^{nab}T_{[2;1]}^{abc}\\
&=&\sum_a(2T^{ama}-T^{maa}-T^{aam}).
\end{eqnarray*}
where $T_{[2;1]}^{abc}=T^{abc}+T^{cba}-T^{bac}-T^{cab}$ means anti-symmetrizing indices $\{12\}$ and then symmetrizing indices $[13]$. This symmetry is described by the Young diagram Fig.\ref{fig: 3site_bases}(b). The bases for $S_t=2$ is
\begin{eqnarray*}
\tilde T_0^{[mn]}&=&\sum_{ab}(\varepsilon^{mab}T_{[2;1]}^{abn}+\varepsilon^{nab}T_{[2;1]}^{abm}).
\end{eqnarray*}
It is easy to check that $\tilde T_0^{[mn]}$ is traceless. 

The representations $S_t=1$ and $S_t=2$ are doubly degenerate because permuting the site indices respects a 2-dimensional representation of the permutation group. The other set of bases can be obtained by replacing $T_{[2;1]}^{abc}$ as $T_{[2;1]'}^{acb}=T^{acb}+T^{cab}-T^{bca}-T^{cba}$ (here $[2;1]'$ means antisymmetrizing $\{13\}$ and then symmetrizing $[12]$):
\begin{eqnarray*}
\tilde V'^m&=&\sum_{nabc}\varepsilon^{mnc}\varepsilon^{nab}T_{[2;1]'}^{acb}\\
&=&\sum_a(2T^{aam}-T^{maa}-T^{ama})\\
\tilde T_0'^{[mn]}&=&\sum_{ab}(\varepsilon^{mab}T_{[2;1]'}^{anb}+\varepsilon^{nab}T_{[2;1]'}^{amb})
\end{eqnarray*}

Fig.\ref{fig: 3site_bases}(c) represents a direct sum of $S_t=1\oplus3$. The bases for $S_t=1$ are given as
\begin{eqnarray*}
\tilde U^{m}&=&\sum_{ab}\delta^{ab}T^{[abm]}
\\&=&{1\over3}\sum_a(T^{maa}+T^{ama}+T^{aam}),
\end{eqnarray*}
and the bases for $S_t=3$ are
\begin{eqnarray*}
\tilde T_0^{[abc]}&=&T^{[abc]}-{1\over5}\sum_m(T^{[amm]}\delta^{bc}+T^{[mbm]}\delta^{ac}+T^{[mmc]}\delta^{ab}).
\end{eqnarray*}

The representation $S_t=1$ occurs three time. It is easy to check that among these $S_t=1$ irreducible representations {\it the bases labeled by different Young diagrams are orthogonal, and the bases labeled by the same Young diagram are linearly independent but not orthogonal. }

We can see that the bases for the irreducible representations are complicated. These bases can be denoted schematically by Young tableaus.

\bibliography{Liuzx}

\end{document}